# PARALLEL COMPUTING FOR THE FINITE ELEMENT METHOD

**C. Vollaire, L. Nicolas** and **A. Nicolas**

CEGELY - UPRESA CNRS 5005 - Ecole Centrale de Lyon-
BP 163 - 69131 Ecully Cedex - France.

vollair@trotek.ec-lyon.fr

laurent@trotek.ec-lyon.fr

nicolas@trotek.ec-lyon.fr

Fax: 04 78 43 37 17

**Abstract:** A finite element method is presented to compute time harmonic microwave fields in three dimensional configurations. Nodal-based finite elements have been coupled with an absorbing boundary condition to solve open boundary problems. This paper describes how the modeling of large devices has been made possible using parallel computation. New algorithms are then proposed to implement this formulation on a cluster of workstations (10 DEC ALPHA 300X) and on a CRAY C98. Analysis of the computation efficiency is performed using simple problems. The electromagnetic scattering of a plane wave by a perfect electric conducting airplane is finally given as example.

**Abstract:** Une formulation de type éléments finis pour la résolution tridimensionnelle de problèmes de diffraction électromagnétique est décrite. Les problèmes ouverts sont modélisés à l'aide d'un couplage avec des conditions aux limites absorbantes. Cet article décrit comment l'utilisation du calcul parallèle a rendu possible la modélisation de grandes structures. De nouveaux algorithmes sont proposés pour une implantation efficace sur une ferme de stations ainsi que sur un CRAY C98. L'analyse des performances est réalisée sur des exemples simples. La réponse électromagnétique d'un avion illuminé par une onde plane est ensuite présentée à titre d'exemple de géométrie réaliste.

**1 INTRODUCTION**

Nodal-based *finite element* (FE) method has been previously developed for microwave problems [1]. Open boundary domains are modeled by coupling with *Absorbing boundary conditions* (ABC). The time harmonic formulation is written in terms of vector fields. Because nodal-based finite elements are used, a penalty term is added to the formulation in order to avoid spurious reflections.

This code has been first developed on scalar workstation. This implies that only simple problems can be modeled. For instance, a 3 wavelengths side cubic geometry (30 cm at 3 GHz), meshed with 10 nodes per wavelength, leads to 81000 complex unknowns. With first order hexahedral FE, 181 Mbytes of memory are necessary to store the matrix. Obviously, the modeling of large geometry such as airplane can not be performed on scalar computers. Only parallel computation actually enables to modelize such devices: it reduces the computation time and, aboveall, it arranges enough memory.

The objective of this paper is to describe how the existing code has been modified in order to implement it efficiently on parallel computers. A cluster of ten DEC ALPHA workstations linked by a FDDI ring has been first used. This distributed memory computer is a *multi instruction - multi data streams* (MIMD) type. *Parallel virtual machine* (PVM) software is used to pass messages between workstations. The formulation has also been implemented on a CRAY C98 which is a shared memory computer (MIMD type) with vector capabilities.

Note that the algorithms and the methodology presented in this paper are not specific to a high frequency electromagnetic formulation. They can actually be applied to any physical problem discretized with a FE method.

**2 FE FORMULATION COUPLED WITH ABC**

We are dealing with frequency domain open boundary electromagnetic field problems. According to Maxwell equations, the magnetic field **H** and the electric field **E** satisfy to the vector wave equations. Following developments are made only for the **H** field formulation. All the steps can be applied to the **E** field formulation as well.

The weak Galerkin formulation of the vector wave equation for **H** is given by (1):

$$\int_v \left[ (\nabla W \times \frac{1}{\varepsilon_r} \nabla \times \mathbf{H}) - W k_0^2 \mu_r \mathbf{H} \right] dv - \int_s \left[ \mathbf{n} \times (W \frac{1}{\varepsilon_r} \nabla \times \mathbf{H}) \right] ds = 0 \qquad (1)$$

A penalty term is added to the formulation to avoid spurious reflections [1]. It makes the field divergence free. Its Galerkin form is given by (2):

$$-\int_v \left[ \nabla W (\nabla \cdot \mathbf{H}) \right] dv + \int_s \left[ W \mathbf{n} \, \nabla \cdot \mathbf{H} \right] ds \qquad (2)$$

ABC is used to truncate the 3D finite element region and to minimize the spurious reflections due to the outer boundary of the FE domain. A 3D vector Engquist-Majda condition [2] is used. Outer surfaces are then rectangles, which is less mesh-consuming for a large number of geometries:

$$\mathbf{n} \times \nabla \times \mathbf{H} \cong g_{ABC}(\mathbf{H}) = j k_0 \mathbf{H}_t - \frac{j}{2 k_0} \nabla_t^2 \mathbf{H}_t \qquad (3)$$

Finally, the formulation for scattering problems in term of total field is given by (4). **H** and $\mathbf{H}_i$ are respectively the total field and the incident field. Two types of surface are considered:

- external surface (*Sext*)

- Perfect Electric Conductor (PEC) surface (*Spec*).

$$\int_v \left[ (\nabla W \times \frac{1}{\varepsilon_r} \nabla \times \mathbf{H}) + W k_0^2 \mu_r \mathbf{H} \right] dv - \int_v \left[ (\nabla W)(\nabla \cdot \mathbf{H}) \right] dv + \int_{Sext+Spec} W \mathbf{n} \nabla \cdot \mathbf{H} ds \\ - \int_{Sext} W g_{ABC}(\mathbf{H}) ds = \int_{Sext} W \left[ g_{ABC}(\mathbf{H}_i) - \mathbf{n} \times \nabla \times \mathbf{H}_i \right] ds \quad (4)$$

## 3 IMPLEMENTATION ON THE CLUSTER OF WORKSTATIONS

Algorithms designed to operate on scalar calculators do not fit to parallel distributed memory computers because data are interdependent. It is not sufficient to add directives of messages passing in a program to allow good performances. Algorithms have to be deeply modified. The key point to respect consists in minimizing the message passing because this step is expensive in CPU time.

The implementation of the formulation on such a computer has been made by using an algorithm to distribute the data over the different workstations. The developed code is a Single Program Multi Data type (SPMD).

Main characteristics of the cluster are :

-10 workstations

-80 Mflops (linpack) per processor

-64 Mo of RAM

-175 MHz clock frequency

-1 Go of swap per processor

-FDDI network in ring (100 Mbits / s)

-PVM 3.3

Performances are analyzed with the PARAGRAPH software. Because its utilization involves the manipulation of post-mortem files, parallel performances can be analyzed only with simple problems. A 60000 degrees of freedom problem is then used because it is the bigger matrix which can be solved on one processor (this step is necessary to evaluate the speedup).

**3.1. Parallel FE algorithm on the cluster of workstations**

3.1.1. Distributing the data

The entire data file is duplicated on each processor. Indeed, the program is a SPMD type performing a parallel reading. An other strategy would consist in storing no more than what is required for one processor at a time and in reading and updating data files during the processing. But this method would result in unacceptable Input/Output (I/O) overhead [3]-[5].

3.1.2. Assembling in parallel

Most of the parallel FE codes use domain decomposition techniques to assemble the FE matrix. These methods are efficient but they require a pre-processing step. Moreover, the number of processors is limited by the decomposition method. This prevents the use of massively parallel computer.

Hence, we have preferred to perform the assembling step of the global matrix by degree of freedom [6], [9], instead of by elementary contribution as in a classical sequential code. It allows to assemble at once the three lines (three coordinates) corresponding to a node. Each element including the considerate node is sought and contributions with the related nodes are computed. These three lines are then compressed by storing only non-zero values. Space is left in the compressed storage if the volume element includes a surface element. Same method is used to introduce ABC on the external boundary and BC on conductors.

This method does not require the creation of the structure of the matrix (called symbolic assembling). To solve a problem meshed with N nodes, if P processors of the cluster are available, each workstation assembles a part of the system including N/P nodes. The load balancing is nearly perfect because of the constant bandwidth of the global matrix. The program is a SPMD type: no message passing is required. Hence, the speedup of the assembling stage is optimal (Fig. 3).

3.1.3. BC on symmetry planes and conductors

Messages passing are necessary to introduce the BC on conductors (only when solving for the **E** field) and on the symmetry planes because a global modification of the matrix system is required. This is due to the method used to introduce the BC (Fig. 1). In the case of the **H** formulation the field verifies (6a) or (6b) on a symmetry plane:

On a symmetry plane: $\mathbf{n} \cdot \mathbf{H} = 0$ (6a)

On an antisymmetry plane: $\mathbf{n} \times \mathbf{H} = 0$ (6b)

The speedup related to this step is low due to the large amount of massage passing (Fig. 3). However, its effect on the global performances of the code is negligible: the length of this step is less than one percent of the total computation time.

3.1.4. Symmetrizing the global matrix

The introduction of ABC and penalty function leads to a non-symmetric system matrix. This one is approximately symmetrized by adding it to its transposed matrix. This has been shown previously to give correct results. This operation requires a large amount of messages passing (Fig. 2). The speedup related to this operation is also low

(Fig. 3). But once again, the time needed to perform this step is small and does not have any significant effect on the total computation time.

3.1.5. Solving in parallel

Because the FE matrix is sparse, iterative methods are used to solve the matrix system. Conjugate Gradient (CG) with diagonal, incomplete Cholesky or block incomplete Cholesky preconditioning methods have then been implemented [7], [8].

In parallel FE codes which use domain decomposition techniques to assemble the FE matrix, the solving step is first performed on the sub-domains and then on the global FE matrix [9]-[12]. With our assembling method (degrees of freedom), the solver works on the entire FE matrix. So, we have chosen to use a small-scale parallelism.

3.1.5.1. Diagonal Preconditioning (DP)

Each processor computes his part of the preconditioning matrix by inversion of the diagonal terms of the FE matrix. No message passing is required and the memory space to store the preconditioning matrix is small. However, the generated preconditioning is low.

The preconditioning is performed by a multiplication vector-vector. A multiplication matrix-vector is also required to compute the residual vector at each iteration. This multiplication is done in parallel (fig. 4). Each processor computes a partial residual vector by multiplicating his part of the FE matrix by the vector duplicated on each processor. The load balancing is nearly perfect because of the constant bandwidth of the matrix.

To operate the concatenation of these partial residual vectors, each processor *broadcasts* it to all the others. Only non-zero values are sent. Then, each processor adds these partial residual vectors to obtain the final residue: this is the SPMD mode. This operation needs ($P^2$-P) messages passing per iteration. An other way to calculate the residual vector consists in sending each partial residual vector to one processor called the master. This one (Master processor number 1) computes the final residue and *broadcasts* it. This Master-Slave (MS) method minimizes the number of communications: only (P-1)x2 messages per iterations are required. On the other hand, it introduces an additional idle time. Furthermore, because only non zero terms are sent, the messages broadcasted by the master are larger than in the SPMD mode: their size is equal to the size of the matrix. For these reasons the SPMD mode is more efficient. In any case, for both methods, the cost of communications is very penalizing, because small-scale parallelism is not adapted to distributed memory computer fitted with this kind of network. Some others methods can be found in [16] but they require more memory space.

Figure 5 shows an average of the state of every processor, on the total execution time, for both methods of concatenation (solving of a 60000 degrees of freedom problem on 4 processors). Figure 6 shows the corresponding speedups.

3.1.5.2. Incomplete Cholesky Preconditioning (ICP)

This preconditioning is performed by the decomposition of the FE in two matrices: A= L . $L^t$ (L and $L^t$ are computed). The building of the incomplete Cholesky matrix is performed by column (5) [15], [16] (fig. 7). This algorithm is implicitly parallel because the L(ij) terms can be computed independently once the diagonal L(jj) term has been computed. The knowledge of both lines j and i is required to compute L(ij). If both lines are not stored on the same processor, a message passing is necessary. Fig. 7 and fig. 8 illustrate this strategy on an example of a 4x4 matrix stored on 3 processors. The incomplete Cholesky matrix is assembled in 5 parallel steps. Between each step, a processors synchronization is required.

For j = 1 to number of lines

$$L_{jj} = \sqrt{A_{jj} - \sum_{k=1}^{j-1}(L_{jk})^2}$$

For i = j+1 to number of lines

$$L_{ij} = \frac{1}{L_{ii}}(A_{ji} - \sum_{k=1}^{i-1} L_{ik} L_{jk})$$

End

End

(5)

While solving using CG, both matrix-vector multiplication and back-forward substitution steps are necessary (6). They have also to be parallelized if possible.

The system A x = b is classically substituted by $(L.L^t)$ x = b. Then, L y = b and $(L^t)$ x = y are computed to solve the system. This algorithm is implicitly sequential because there is a back dependency on y and x: for example, the first processor computes his part of y and broadcasts it to all the others (in SPMD mode). The second processor can then start to compute and so on. This step is very penalizing in term of parallel performances, because it is performed at each iteration of the solver (fig. 9). Figure 10 shows the corresponding speedup.

**Forward substitution**

For i = 1 to number of lines (n)

$$y_i = (b_i - \sum_{k=1}^{k=i-1} L_{ik}\, y_k)/L_{ii}$$

End

(6)

**Back substitution**

For i = number of lines (n) to 1

$$xi = (yi - \sum_{k=i+1}^{k=n} Lki\, xk)/Lii$$

End

This method allows to reduce the number of iterations necessary to solve the system of equations (tab. 1). However, because of the large amount of message passing required, it cannot be applied to large problems. The memory space needed to achieve the ICP is 1.5 time larger than when using the DP because L and $L^t$ have the same structure than A after its symmetrization. $L^t$ is built to achieve a quick access by columns to the terms. This means that the terms are stored by lines in L and by columns in $L^t$.

3.1.5.3. Block Incomplete Cholesky Preconditioning (BICP)

To avoid messages passing during the building of the incomplete Cholesky matrix, this one is assembled only with the terms stored on the considerate processor. So a part of the matrix is not built (fig. 11), and the terms effectively assembled are approximated because of the back dependency. This method applied with one processor corresponds to the classical ICP.

Compared to the Incomplete Cholesky Preconditioning, this scheme leads to a degradation (tab. 1) of the preconditioning. On the other hand, the preconditioning matrix is now constituted by independent sub-blocks. Each processor can compute independently his part of the result vector. The concatenation of these partial results, necessary at the end of the forward and the back substitution, is performed by message passing in SPMD mode. As show is tab. 1, the number of iterations required to solve the system of equations depends on the number of processors available: the increase of the number of processors leads to an increase of the number of iterations. On the other hand, the parallel rate of the method is increased (fig. 9-12). Figure 12 shows that the processors are better used than when solving with a classical ICP. The number of operations required to compute both preconditioning matrix and result vector depends on the number of processors available because of the not-computed terms in the preconditioning matrix. So, the parallel performances cannot be estimated in term of speedup because the solver on 2 processors is three times faster than on 1 processor. Figure 13 shows the CPU time per processor for a solving of a 60000 degrees of freedom matrix versus the number of processors used.

For this size of matrix, the method loses efficiency when more than 5 processors are used. The increase of the matrix size tends to reject this limit.

Note that the memory space needed to achieve the BICP is 1.5 time larger than when using the DP because L and $L^t$ have the same structure than A after its symmetrization.

3.1.5.4. Comparison between methods

Table 1 compares the performances of the preconditioning methods when solving a 60000 degrees of freedom matrix on 2, 4 and 8 processors.

For this small example, on 2 processors, the Diagonal Preconditioning is the most efficient. The Block Incomplete Cholesky Preconditioning becomes interesting when more than 4 processors are used. For large problems, because of the large amount of messages passing, the building of the incomplete preconditioning matrix requires too much CPU time. Only the Block Incomplete Cholesky Preconditioning can be used.

3.2. Modeling a large problem

The electromagnetic scattering of a plane wave by a perfect electric conducting airplane (fig. 14) is presented as realistic problem. The frequency of the incident plane wave is 0.3 Ghz and this problem is meshed with 51183 nodes (307098 degrees of freedom). To solve the system of equations of this problem on 8 processors, the conjugate gradient with Diagonal Preconditioning requires 7476 iterations and 80280 s per processor while the conjugate gradient with the Block Incomplete Cholesky Preconditioning requires only 4902 iterations and 58950 s per processor.

**4 IMPLEMENTATION ON THE CRAY C98**

The CRAY C98* is a high performance parallel vector computer. It is a shared memory *multi instruction, multi data streams* (MIMD) computer. It supports therefore a parallelism of strong scale. Indeed, every processor can execute a different program or a part of same program. All the processors have access to the global shared memory through a central connection.

Main characteristics of the CRAY C98 are the following:

-8 vector processors

-1 GFlops of peak performance per processor

-4 GBytes memory (= 512 Mw)

-8 128-words vector registers

-4,17 ns clock cycle

-120 GBytes of disk space

**4.1. Programming model**

Parallelism and vectorization are introduced by compiler directives. A vectorization directive leads the compiler to use the vector registers. There are several ways to use parallelism. Two of them are mainly used:

* This work was supported in part by the Institut du Développement et des Ressources en Informatique Scientifique (CNRS).

-The first one consists in putting a parallelization directive before a loop. This one will be split, and the number of iterations performed by each processor can be specified. The variables used in the loop must have a scope, which means that they must be either private or shared. Each processor has its own copy of the private variable, while a shared variable is shared by all processors. Memory write conflicts on a shared variable may result in data corruption and must be avoided.

-The second way consists in creating a parallel region, which includes a loop. All the variables inside the parallel region must have a scope. The loop is parallelized, but the private variables can be used outside of the loop. This is useful for initializations, or to create private buffers to avoid memory conflicts. Semaphores can also prevent these conflicts.

A loop will be broken in packets that are long enough to enable vectorization (128 Words). Data streams must be as longer as possible to allow a good vectorization.

The CRAY compiler can try to parallelize and vectorize the code automatically due to compilation directives, by checking the data dependencies in the loops. It is automatic, but it works only on simple loops. For example, a subroutine call inside a loop prevents it from being parallelized. This automatic method leads then to bad parallel performances.

In order to parallelize loops which have not been made automatically, an higher level tool (BROUSE) may be used. It tries to find the scope of each variable in a specified loop. There are also mainly profiling tools as 'hpm', 'proview', 'atexpert', ... which give to the user many information about the performances of the code such as: parallelism ratio, average vector length, parallel / sequential portion of the code, most important loops / subroutines, ...

### 4.2. FE algorithm on the CRAY

The algorithms operating on mono-processor calculator do not need to be entirely revised in order to work on the CRAY C98. The code remains the same and runs quite immediately with more or less automatic parallelism-vectorization and progressively gains parallelism-vectorization while the user manually adds parallelization-vectorization directives.

Our code was first developed on a workstation, in a well structured way with many subroutine calls, error handling, .... Those points prevent the compiler from getting a good parallelization level. The compiler itself got only 17% of parallelization on a problem meshed with 10000 nodes. Only simple loops, like initializations had been processed. Neither the matrix assembly portion nor the solver had gained parallelism. They only gained some vectorization. It was then necessary to help the compiler to get better results by adding manually compiler directives and by finding the scope of every variables.

The experience in the implementation of codes on such computer has shown that the matrix representation has a great influence on vector performances. So, two types of representations have been used.

*A. Sparse row-wise matrix representation (storage # 1)*: because of the FE matrix is sparse and symmetric, only the non zero terms of its lower part are stored [2] after its symmetrization.

*B. Redundant Sparse row-wise matrix representation (storage # 2)* : because the algorithms used in the solver need to access to the FE terms by column, the entire FE matrix stays in memory even after its symmetrization. The memory space used is twice larger, but the access to the non zero terms of a column are adjacent in memory [16].

Some other methods like Sky line matrix representation have been implemented. Because all the terms contained in the bandwidth of the FE matrix are stored, the number of terms stored for a line of the FE matrix increases with the size of the problem. This is due to the renumbering algorithm. So, this method requires too much memory and it can not be used to compute realistic devices.

All the performances are analyzed as previously with a 10000 nodes problem because the use of the ATEXPERT software (analysis of parallel performances) involves a mono-processor computation.

4.2.1. Creation of the global matrix structure

This step prepares the storage of the non-zero terms of the matrix. Its parallelization is difficult because there are many data dependencies. This part of code is especially vectorized.

4.2.2. Assembling by elementary contributions

This step is easy to parallelize. The elementary matrix related to a FE needs to have a private scope, and semaphores are used when the global matrix, stored in shared memory, is modified. So a parallel region is created which includes the loop on the FE. The global matrix is modified in a critical region, inside the parallel region, in which the code is executed in a sequential way.

An other solution can be found in [4] but it needs a 'pre-processing' step to split the set of FE into subsets made of independent elements. This method is efficient when write access conflicts in the global matrix for contiguous elements decrease the performances.

4.2.3. Introduction of BC on conductors (E field solving) and on the symmetry planes

Due to the method used to introduce BC, a global modification of the matrix system is required. Each processor performs a part of the modifications which are made on the global matrix. Semaphores are used to avoid memory conflicts.

4.2.4. Symmetrization of the matrix

Non-symmetric global matrix is symmetrized by adding it to its transposed matrix: all the terms of each line are added with those corresponding of the transposed matrix. The external loop (on the lines) is parallelized while the internal one (on the columns) is vectorized. The variables used for this step can be shared because no memory conflict is possible.

Figure 15 shows the speedups for the creation of the FE matrix (steps 4.2.1, 4.2.2, 4.2.3, 4.2.4) versus the data representation. The creation of the global structure matrix decreases the parallel performances because this step is especially sequential.

4.2.5. Solver

As previously presented, the conjugate gradient is used to solve the system of equations.

4.2.5.1. Diagonal Preconditioning

The use of this method requires only a multiplication matrix-vector per iteration to compute the new residual vector. So this step is parallelized by splitting the loop on the lines.

For the storage # 1, each processor computes a partial residual vector in private memory (parallel region). The addition of these partial results is performed in shared memory in a critical region (vectorization).

For the storage # 2, the multiplication is done in shared memory because the entire FE matrix is stored.

The speedups versus the matrix representation of the CG with the Diagonal Preconditioning are presented in figure 16.

4.2.5.2. Incomplete Cholesky Preconditioning

The storage # 1 requires the research of the lines (i) with a non zero term on the column (j). The parallelization is made by splitting this loop. Indeed, this algorithm needs the multiplication of the line (j) by the line (i). So an other search on the terms of the line (i) is necessary for all the terms of the line (j). This matrix representation is not adapted to the Incomplete Cholesky Preconditioning because of the vector performances obtained (tab. 2 and 3). On the other hand, only the lower part of the preconditioning matrix is built. Because only the lower part of the FE is stored in memory after its symmetrization, no more allocation of memory is required: the preconditioning matrix is stored in memory at the place of the upper FE matrix.

The storage # 2 allows to access directly to the lines (i) which have a non zero term on the column (j). Therefore, this loop is parallelized. The multiplication of the line (j) by the line (i) requires a search (for all the terms of the line (j)) on the terms of the line (i). This matrix representation improves the vector performances of the code during the building of the preconditioning matrix (tab. 2 and 3). The entire preconditioning matrix is built to keep the advantage of the adjacent access to the terms of a same column. Because the entire FE matrix is stored in memory after its symmetrization, this type of representation requires to double the memory space.

Figure 17 shows the speedups obtained for the building of the Cholesky matrix versus the matrix representation.

The parallelization of the forward substitution cannot be made by splitting the loop on the line because of the back dependency. So, the produce of the terms L(ik) . y(k) is parallelized by creating a parallel region. The addition of the partial results is made in a critical region. The access to the terms L(ik) is performed by line, leading to good vector performances.

The parallelization of the back substitution is made in the same way but the access to the terms L(ki) decreases the vector performances (tab. 2 and 3). So the influence of the type of representation is the same as in the building of the Cholesky matrix. Figure 18 shows the speedups for back-forward substitution versus the matrix representation.

For the storage # 1, the research of the term L(ki) is penalizing in term of vector performances (tab. 2 and 3). The overhead introduced by splitting the loop on the produce L(ki) . x(k) is then negligible.

The use of the storage # 2 involves less efficiency in term of parallel performances because of the overhead introduced by splitting the produce L(ki) . x(k). On the other hand, because the terms L(ki) are adjacent in memory, the vectors performances are increased (tab. 2 and 3).

**4.2. Comparison between the methods**

Table 2 compares parallel and vector performances of the preconditioning methods when solving a 60000 degrees of freedom matrix with both types of matrix representation (total CPU time). Table 3 compares same performances of the preconditioning methods when solving a 307098 degrees of freedom matrix (scattering by a pec airplane). Both tables show the efficiency of the storage #2.

## 5 CONCLUSION

We have presented in this paper the implementation of an electromagnetic scattering code on parallel shared and distributed memory computers.

### 5.1. Cluster of stations

The parallel performances obtained when building the FE matrix are satisfying.

Because of the low preconditioning, the cost of communications with the Diagonal Preconditioning is penalizing. This is due to the small-scale parallelism which is not adapted to distributed memory computer equipped with this type of network. The Incomplete Cholesky Preconditioning allows to reduce the number of iterations but cannot be used on large devices because of the large amount of messages passing required. So, our new preconditioning method named 'Block Incomplete Cholesky Preconditioning' seems a good compromise in term of convergence rate and CPU time. However, the memory space needed to achieve the BICP is 1.5 times larger than when using the DP.

Computed examples show that the modeling of realistic problem is still not possible on such a cluster because of the memory available. For example, the illumination of a fighter by a 3 GHz plane wave -high range radar freqency- would lead to a $10^6$ nodes problem.

On an other hand, this code is immediately implementable on a computer such as the CRAY T3E. This parallel distributed memory computer is a MIMD type. Given its characteristics (256 processors and 32 Go of memory) it should allow to arrange enough memory to modelize large devices. Moreover, this machine is equipped with a very high performances network.

### 5.2. CRAY C98

The adaptation to a shared memory computer is easy because it does not ask a complete restructuring of program. The code remains the same and runs quite immediately. The CRAY compiler can try to parallelize and vectorize the code automatically due to compilation directives, by checking the data dependencies in the loops. It is automatic, but it leads to bad parallel performances. So the programmer has to manually add parallelization-vectorization directives and to cheek the scope of every variable to keep the control on the parallelism granulity.

The relative performances due to the vectorization are very low with a classical matrix representation method because the data streams are short (the matrix is sparse). The use of the Redundant Sparse row-wise storage allows to obtain acceptable vector performances but it requires to double the memory space. Our experience shows that it is more important to favor the vectorization than the parallelization because every computation node of this machine is especially vectorial.

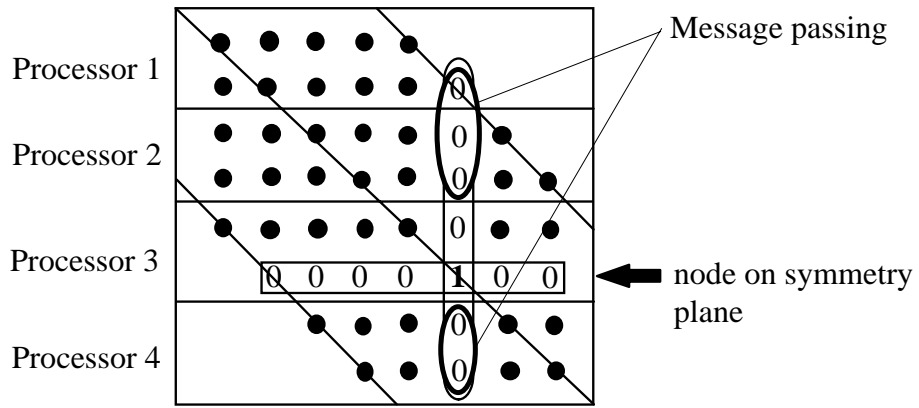

**Figure 1.** Message passing for the symmetry planes - Example of 4 processors.

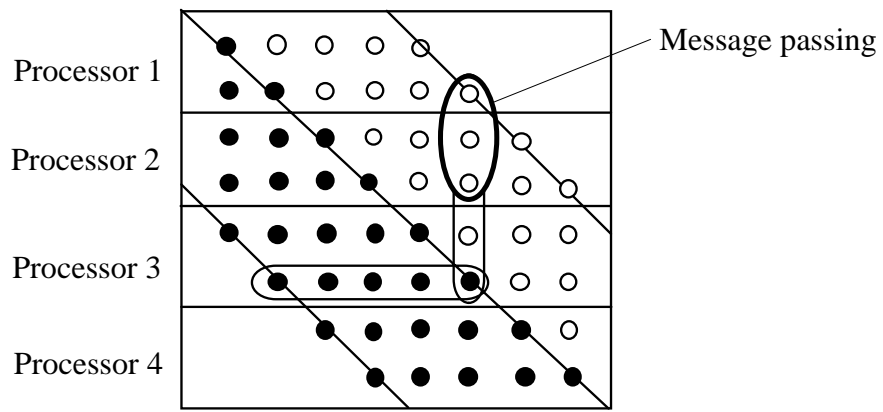

**Figure 2.** Message passing for the symmetrization - Example of 4 processors.

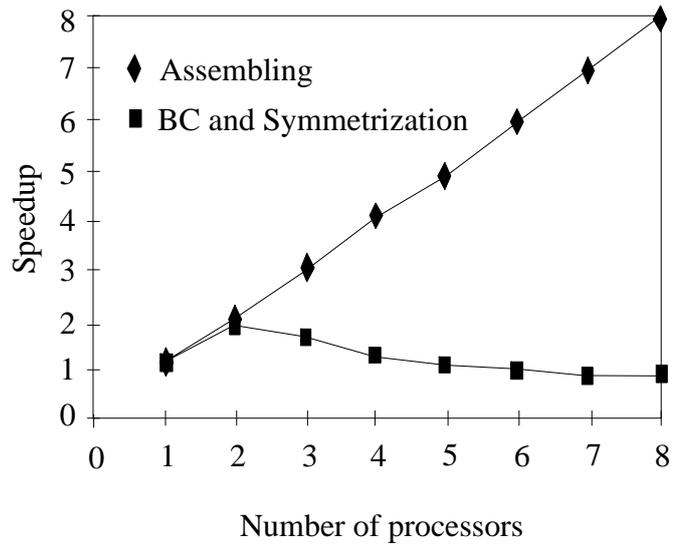

**Figure 3.** Speedups for a 10000 nodes problem (60000 degrees of freedom).

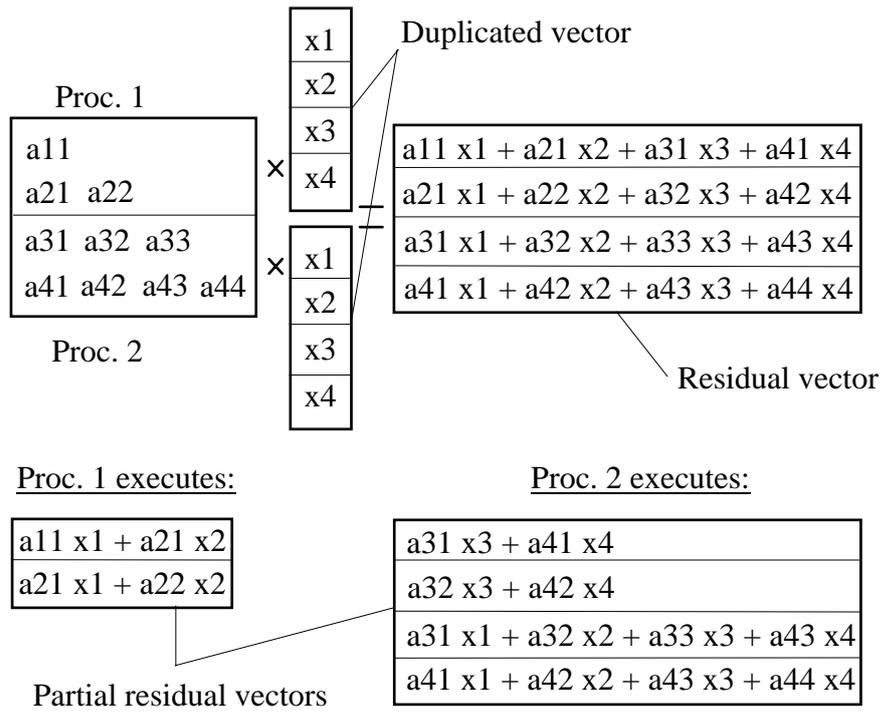

**Figure 4.** Parallel matrix-vector multiplication- Example of a 4x4 matrix on 2 processors.

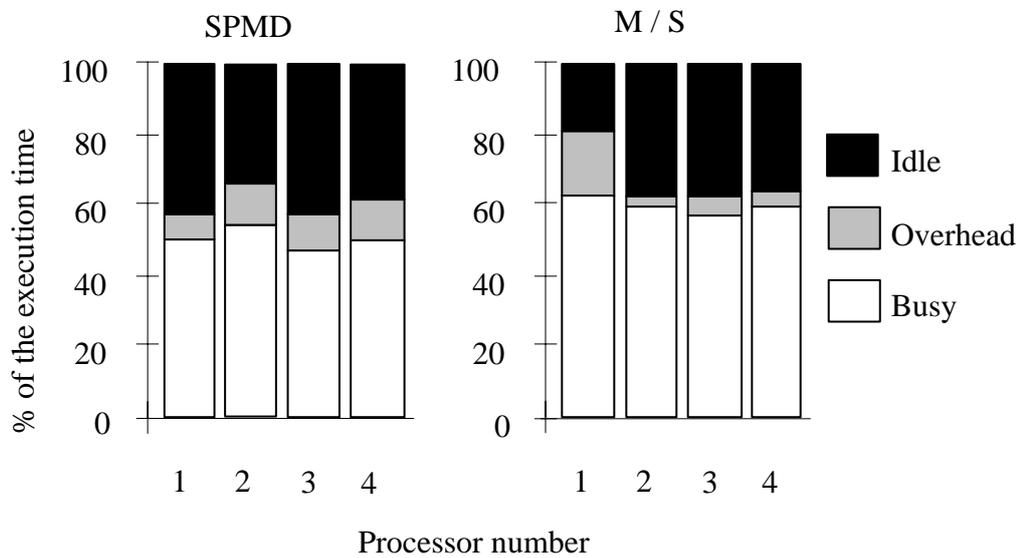

**Figure 5.** Average of the processors states for the CG with the Diagonal Preconditioning.

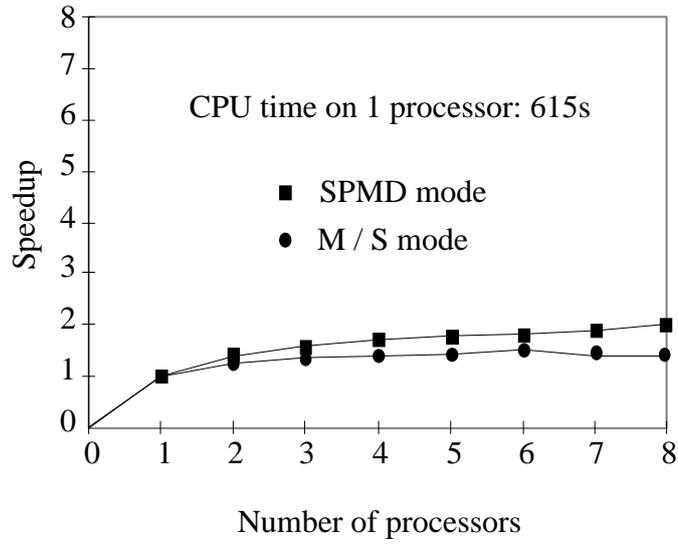

**Figure 6.** Speedups of the CG with the Diagonal Preconditioning for both methods of concatenation.

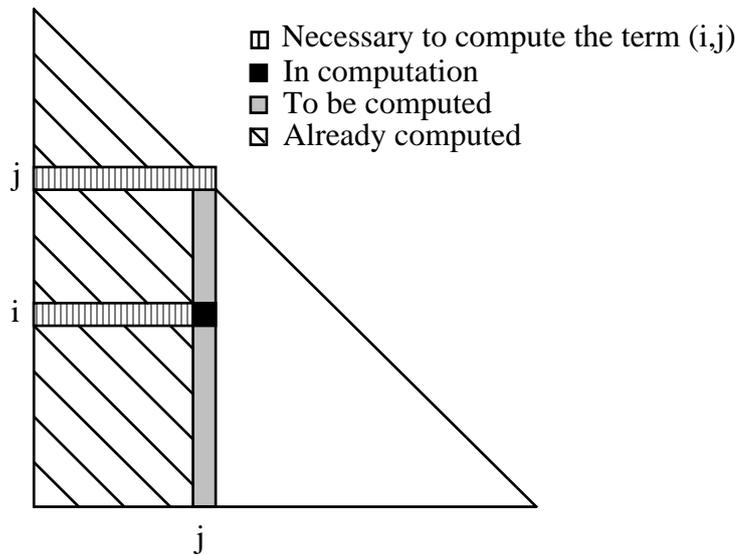

**Figure 7.** Building of the Incomplete Cholesky matrix L by columns.

|        | Proc. 1 | Proc. 2 | Proc. 3 |
|--------|---------|---------|---------|
| Processor 1 | a11 | | |
| | a21  a22 | | |
| Processor 2 | a31  a32  a33 | | |
| Processor 3 | a41  a42  a43  a44 | | |

FE matrix stored on 3 processors

| Step | Proc. 1 | Proc. 2 | Proc. 3 |
|------|---------|---------|---------|
| 1 | Computation of L11 L11T → L11 | | |
| 2 | Computation of L21 L21T<br>Computation of L22 L22T → L21 L22 | Comp. of L31  L31T | Comp. of L41  L41T |
| 3 | ↙ L31 L32 L33 | Comp. of L32  L32T<br>Comp. of L33  L33T | Comp. of L42  L42T<br>↖ L31 L32 L33 |
| 4 | Insertion of L31T  L32T | L41 L42 L43 L44 | Comp. of L43  L43T<br>Comp. of L44  L44T |
| 5 | Insertion of L41T  L42T | Insertion of L43T | |

**Figure 8.** Building of L and L$^t$ in 5 parallel steps.

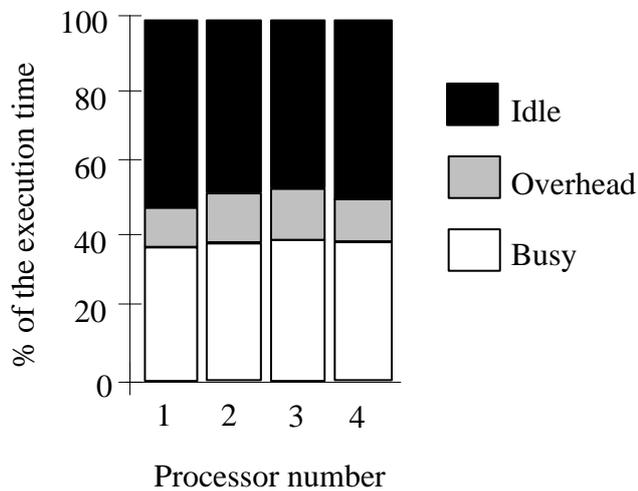

**Figure 9.** Average of the processors states for the CG with the Incomplete Cholesky Preconditioning.

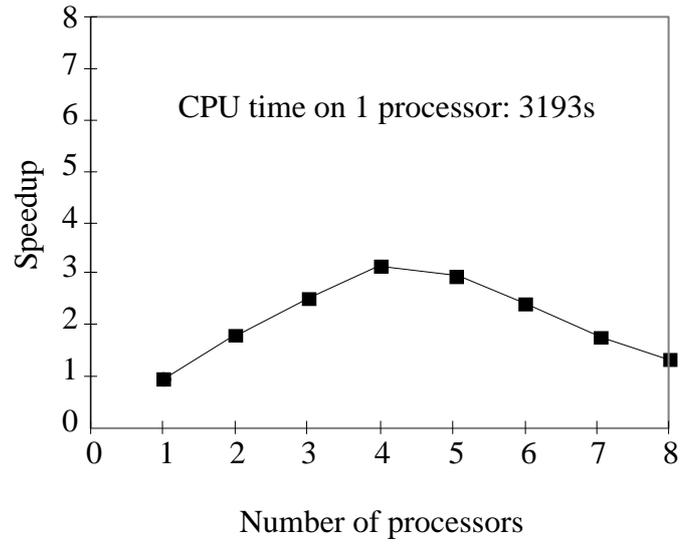

**Figure 10.** Speedup of the CG with the Incomplete Cholesky Preconditioning

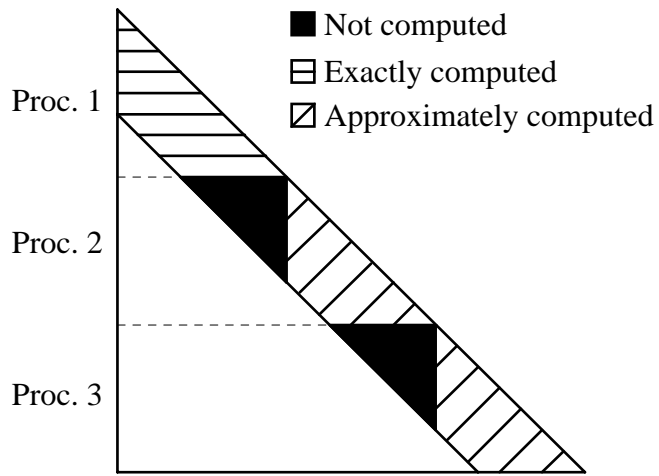

**Figure 11.** Lower triangular matrix L constructed per block (3 processors).

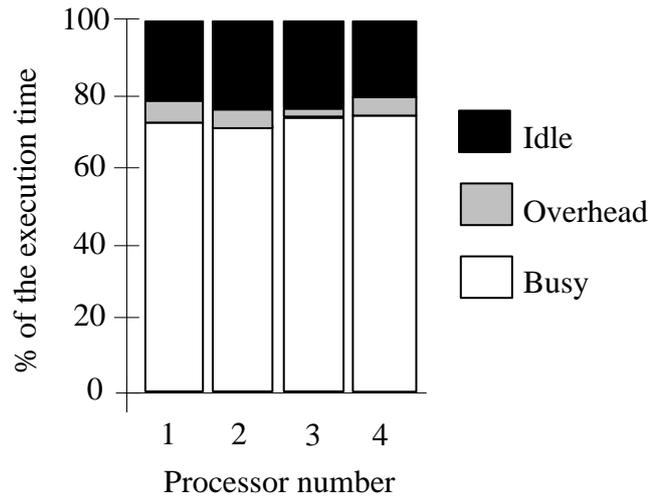

**Figure 12.** Average of the processors states for the CG with the Block Incomplete Cholesky Preconditioning.

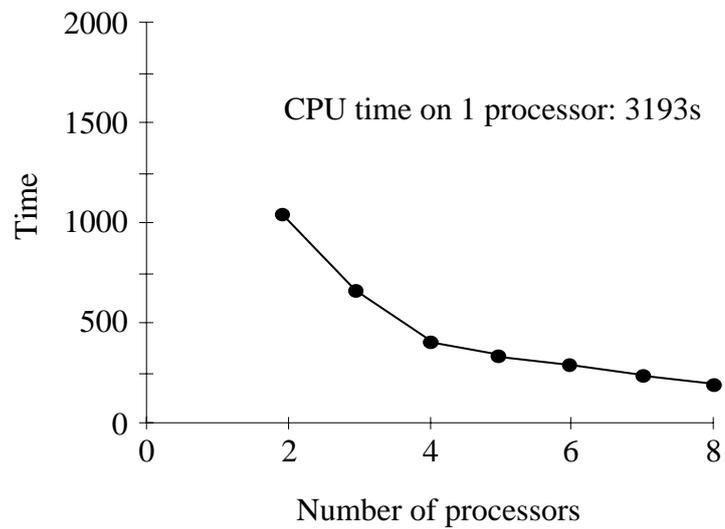

**Figure 13.** CPU time used for the CG with the Block Incomplete Cholesky Preconditioning.

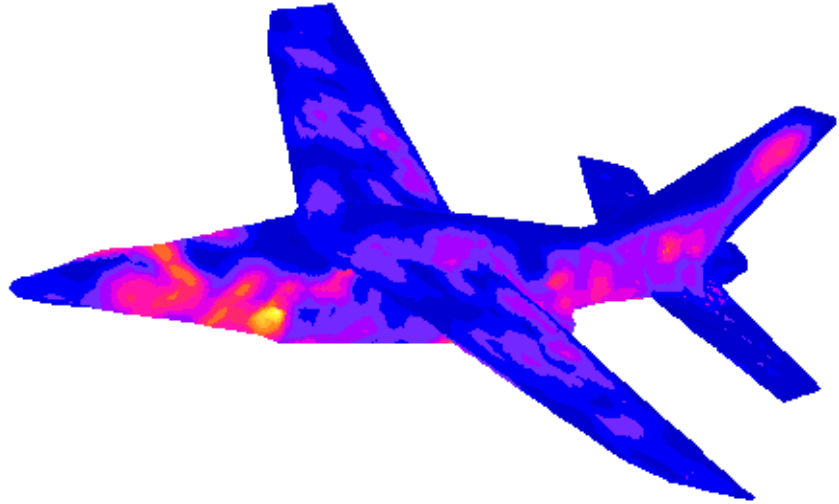

**Figure 14.** Perfect electric conductor aeroplane enlighted by a plane wave (magnitude of the magnetic field **H**).

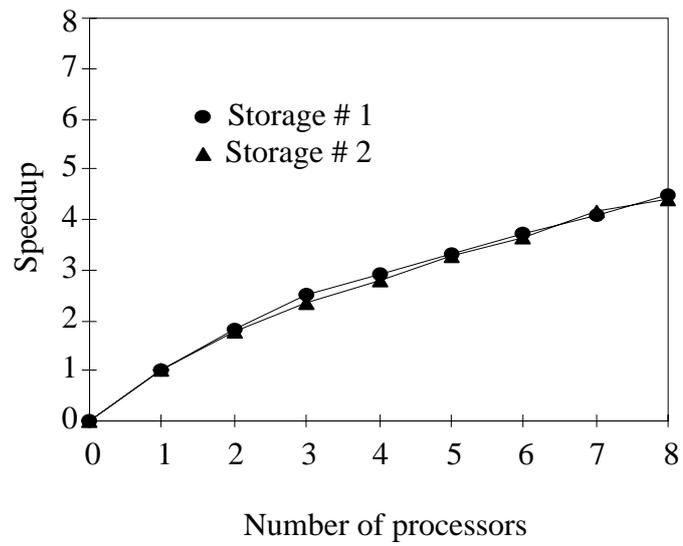

**Figure 15.** Speedups for the assembling step.

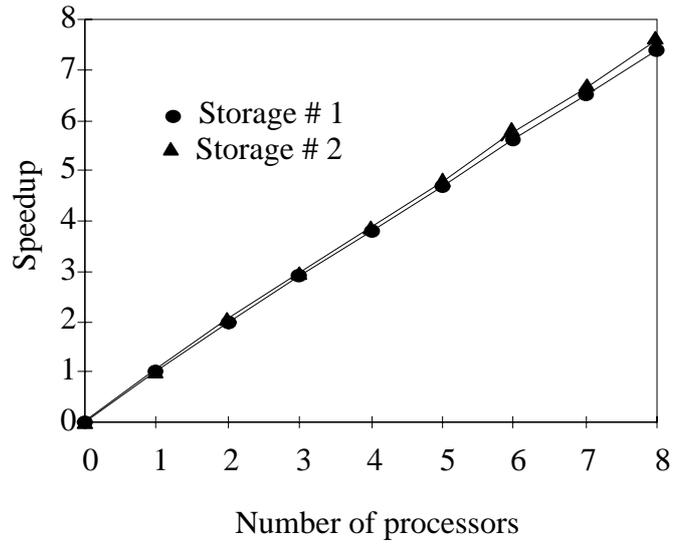

**Figure 16.** Speedups for the CG with the Diagonal Preconditioning.

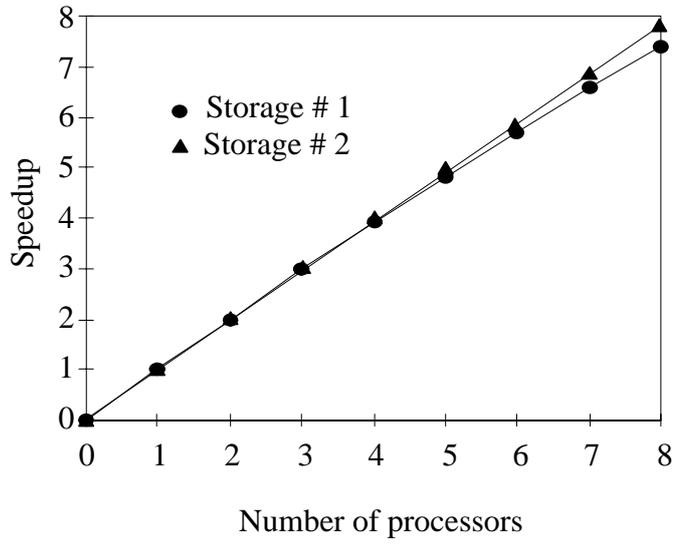

**Figure 17.** Speedups for the building of the Cholesky matrix.

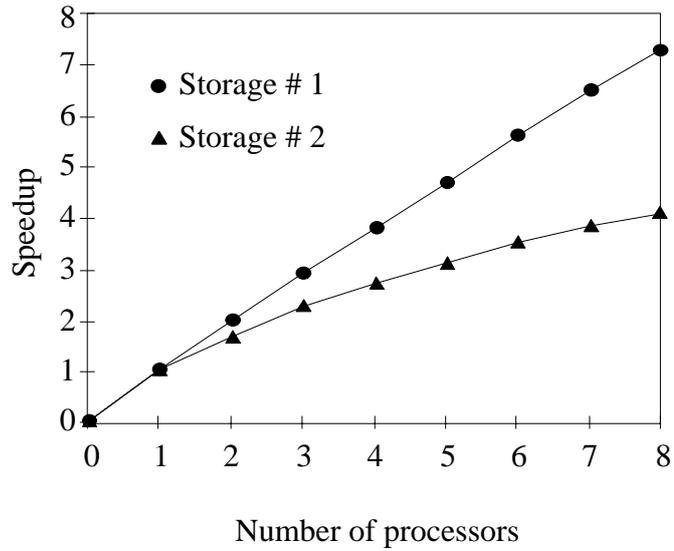

**Figure 18.** Speedups for the back-forward substitution.

| Number of processors | 2 | | 4 | | 8 | |
|---|---|---|---|---|---|---|
| | time (s) | ite. | time (s) | ite. | time (s) | ite |
| Diagonal Precond. | 415 | 204 | 367 | 204 | 316 | 204 |
| Cholesky Precond. | 1781 | 59 | 901 | 59 | 1773 | 59 |
| Block. Cho. Precond. | 1133 | 59 | 353 | 61 | 234 | 81 |

**Table 1.** CPU time per processor and number of iterations versus the preconditioning method and the number of processors.

| | | CPU time (s) | Mflops | Iterations |
|---|---|---|---|---|
| Diag. Precond. | storage # 1 | 263 | 20 | 232 |
| | storage # 2 | 187 | 40.2 | 232 |
| Cho. Precond. | storage # 1 | 5314 | 2.3 | 76 |
| | storage # 2 | 733 | 10 | 76 |

**Table 2.** Total CPU time, vector performances and number of iterations versus the preconditioning method and the matrix representation (solving on 8 processors).

| | | CPU time (s) | Mflops | Iterations |
|---|---|---|---|---|
| Diag. Precond. | storage # 1 | 23035 | 56 | 41507 |
| | storage # 2 | 9981 | 161 | 41507 |
| Cho. Precond. | storage # 1 | too much CPU time consuming | | |
| | storage # 2 | 8072 | 113 | 13667 |

**Table 3.** Total CPU time, vector performances and number of iterations versus the preconditioning method and the matrix representation (solving on 8 processors).